\documentclass{aa}
\usepackage{amssymb}
\usepackage{amsmath}
\usepackage{xspace}
\usepackage{graphicx}
\usepackage{enumerate}   
\usepackage{color,xcolor}
\usepackage{txfonts}

\usepackage{natbib}
\usepackage[colorlinks=true,citecolor=blue,urlcolor=blue,linkcolor=blue]{hyperref}

\def\gr{$\gamma$-ray}

\begin{document}

\title{Constraint on magnetized galactic outflows from LOFAR rotation measure data}

\author{J.Blunier$^{1}$, A.Neronov$^{1,2}$
}
\institute{
    Laboratory of Astrophysics, Ecole Polytechnique Federale de Lausanne, CH-1015, Lausanne, Switzerland
    \and Université de Paris Cité, CNRS, Astroparticule et Cosmologie, F-75013 Paris, France
}

\authorrunning{Blunier \& Neronov}
\titlerunning{Constraint on magnetized outflows from LOFAR}
\abstract
{
Outflows from galaxies, driven by active galactic nuclei and star formation activity, spread magnetic fields into the intergalactic medium. Importance of this process can be assessed using cosmological magneto-hydrodynamic numerical modelling of the baryonic feedback on the Large Scale Structure, like that of Illustris-TNG simulations. We use the Faraday Rotation Measure data of LOFAR Two-Meter Sky Survey (LOTSS) to test the Illustris-TNG baryonic feedback model.  We show that the Illustris-TNG over-predicts the root-mean-square of the residual Rotation Measure in LOTSS. This suggests that the "pollution" of the intergalactic medium by the magnetized outflows from galaxies is less important than the  estimate from the Illustris-TNG. This fact provides a support to the hypothesis that the volume-filling large scale magnetic fields found in the voids of the Large Scale Structure are likely of cosmological origin.  
}

\maketitle

\section{Introduction}

Magnetic fields are present in different elements of the  intergalactic medium, including voids \citep{2010Sci...328...73N,2023A&A...670A.145A} and filaments \citep{2021MNRAS.505.4178V,2023SciA....9E7233V,2023MNRAS.518.2273C}. These fields may originate from the Early Universe \citep{2001PhR...348..163G,2013A&ARv..21...62D,2021RPPh...84g4901V}, or be spread by outflows from galaxies as a part of the baryonic feedback on the Large Scale Structure (LSS) \citep{2006MNRAS.370..319B}. The cosmological magnetic fields may potentially be found in their original form in the voids of the LSS. However, it is not clear a-priori, if the baryonic feedback fields are contained in magnetized bubbles around galaxies or if they can overflow into the voids. If this happens, the strength of the baryonic feedback field in the voids may happen to be larger than that of the cosmological field. This would make measurement of the parameters of the cosmological magnetic field impossible. 

The state-of-art modelling of the baryonic feedback process is provided by the cosmological Magneto-Hydro-Dynamical (MHD) simulations like those of Illustris-TNG \citep{2018MNRAS.480.5113M}. It indicates that the strength of the baryonic outflows from galaxies is probably not sufficient for spreading the magnetic fields into the voids and that the feedback fields stay contained within magnetized bubbles around galaxies \citep{2023MNRAS.519.4030A}. However, the feedback model has many parameters that are only partially constrained by galaxy-scale observables, so that the model estimates of the parameters of magnetic fields spread by the galactic outflows still suffer from large uncertainties. 

Faraday Rotation Measure (RM) of extragalactic sources is sensitive to the parameters of the baryonic feedback fields \citep{2023MNRAS.519.4030A} and thus the RM data can be used to reduce the uncertainty of the baryonic feedback modelling. In what follows we use the RM dataset of the Data Release 2 of the LOFAR Two-Meter Sky Survey (LOTSS) \citep{2023MNRAS.519.5723O,2023MNRAS.518.2273C} to constrain the magnetic fields spread by the baryonic outflows in the intergalactic medium. We compare the Residual Rotation Measure (RRM) derived from the LOTSS dataset by subtracting the estimated Galactic component of the RM with predictions of the Illustris-TNG model, following the methodology of \cite{2023MNRAS.518.2273C} for the formulation of model predictions for subsequent comparison with the LOTSS data.

\section{Faraday rotation measure of extragalactic sources}

The rotation of the polarisation angle $\psi$ of a radio signal at the wavelength $\lambda_{obs}$ coming from a source at a distance $l_s$ is described by the  Faraday Rotation Measure (RM) 
\begin{equation}
\label{eq:RM}
    RM=\frac{\psi}{\lambda^2_{obs}}=\frac{e^3}{2\pi m_e^2 c^4}
    \int_0^{l_s}n_e(l)B_{||}(l)\left(\frac{\lambda(l)}{\lambda_{obs}}\right)^2dl
\end{equation}
where $n_e$ is the proper electron density, $B_{||}$ is the proper component of magnetic field parallel to the line-of-sight, $\lambda(l)$ is the wavelength of the signal at the distance $l$ and $m_e,c$ are the electron mass and the speed of light. The co-moving distance scales with the redshift $z$ as
\begin{equation}
\frac{dl}{dz}=\frac{c}{H_0(1+z)\sqrt{\Omega_M(1+z)^3+\Omega_\Lambda}}
\end{equation}
where $H_0\simeq  67.74$~km/(s Mpc) is the Hubble constant, $\Omega_m \simeq 0.3089$ and $\Omega_\Lambda \simeq 0.6911$ are the present-day matter and cosmological constant fractional densities \citep{2016A&A...594A..13P}.

%\textcolor{red}{
%Jeffrey, your definition is 
%\begin{eqnarray}
%    &&\frac{dl}{dz}\simeq \frac{1}{(1+z)}\left(\frac{c}{H}-%\frac{\tilde l}{(1+z)}\right)\simeq \frac{1}{(1+z)}\left(\frac{c}{H}-\frac{cz}{H(1+z)}\right)\nonumber\\
%    && \simeq \frac{1}{(1+z)}\left(\frac{c(1+z)-cz}{H(1+z)}\right)\simeq \frac{c}{(1+z)^2H_0\sqrt{\Omega_M(1+z)^3+\Omega_\Lambda}}\nonumber
%\end{eqnarray}
%(please check), so overall there is only one factor of $(1+z)$ lost, for small $z$ in Eq. (3), right?
%}
Introducing co-moving magnetic field strength $\tilde B_{||}=B_{||}/(1+z)^2$ and electron density $\tilde n_e=n_e/(1+z)^3$, one can rewrite Eq. (\ref{eq:RM}) as 
\begin{equation}
\label{eq:RM1}
    RM=\frac{e^3}{2\pi m_e^2 c^3 H_0}\int_0^{z_s}
    \frac{(1+z)^2\tilde n_e\tilde B_{||}}{\sqrt{\Omega_M(1+z)^3+\Omega_\Lambda}}dz
\end{equation}

The electron densities and magnetic field $\tilde n_e, \tilde B_{||}$ found in the right-hand side of the above equation can be extracted from cosmological simulations to get predictions for the dependence of the extragalactic RM on the redshift. We do this for the Illustris-TNG simulations. 

\section{Illustris-TNG model}

The Illustris-TNG (The Next Generation) project presents a suite of large-scale MHD cosmological simulations with different volume sizes \citep{2018MNRAS.475..676S,2018MNRAS.475..648P,2018MNRAS.477.1206N,2018MNRAS.480.5113M,2018MNRAS.475..624N} that aims to model the formation and evolution of galaxies in the Universe. 

We focus on the simulation box of TNG100 at its highest resolution. The TNG100 simulation was run using the moving-mesh code AREPO \citep{2010MNRAS.401..791S} to solve the 
equations of ideal MHD \citep{2011MNRAS.418.1392P}, with account of gravity. The Illustris-TNG model is designed to reproduce several properties and statistics of observed 
galaxies including the stellar-to-halo mass relation and the black-hole -- stellar mass relation at redshift $z = 0$.  The simulations start from redshift $z = 127$ and stop at redshift $z = 0$. TNG100 is a box of the size $106.5$ comoving Mpc 
(cMpc) with periodic boundary conditions including $2 \times 
1820^3$ particles. TNG100 has a dark matter particle mass of 
$7.5\times 10^6M_\odot$ and a baryonic mass of $1.4\times 10^6M_\odot$. A hierarchical structure is established for every run through 100 snapshots spanning the evolution of the simulation from $z = 127$ to $z = 0$, each containing the characteristics of all particles. We  consider only gas particles, also known as gas cells, with the following properties: the coordinates, the density, the mass, the magnetic field and the electron abundance. 

Illustris-TNG reproduces the typical magnetic field of $1-10\ \mu$G present in galaxies. The galactic magnetic fields are produced in result of compression and action of dynamos on an initial seed homogeneous field with the comoving strength $\tilde B_0=10^{-14}$~G. In the absence of magnetized galactic outflows and dynamos, this primordial magnetic field would be compressed during the structure formation, respecting conservation of the magnetic flux, so that its strength would scale with the density as $\rho^{2/3}$. Deviations from this scaling in moderately overdense regions of the LSS occur due to the magnetized outflows from galaxies driven by star formation and active galactic nuclei (AGN) \citep{2018MNRAS.480.5113M,2023MNRAS.519.4030A}. The phenomenological model of the outflows depends on a range of parameters. For example, the AGN feedback model \citep{2017MNRAS.465.3291W} assumes that  a minimal black hole mass $M_{min}=1.18\times 10^6M_\odot$ is found in the minimum of the gravitational potential of halos as soon as their masses reach $7.38\times 10^{10} M_\odot$. The black hole starts accreting and can enter two different modes depending on the Eddington ratio $f_{Edd} = \dot M_{BH} /\dot M_{Edd}$ where $\dot M_{BH}$ and $\dot M_{Edd}$ are the actual and Eddington accretion rates respectively. When $f_{Edd }> 0.1$,  the energy feedback rate of an outflow generated by the AGN activity is supposed to be  $\dot E_{high}=\epsilon_{f,high}\epsilon_r\dot M_{BH}c^2$, where $\epsilon_{f,high},\epsilon_r$ are the phenomenological efficiencies of converting the black hole accretion power $\dot M_{BH}c^2$ into thermal energy of the outflowing gas and radiation. If $f_{Edd}<0.1$, only kinetic energy is injected into the gas cells surrounding the black hole, with no direct conversion into radiation and/or thermalisation. Most of the parameters of the feedback model cited above are not constrained individually and are instead adjusted based on a limited number of galaxy-scale observables. New observables, like the RM of extragalactic sources, discussed below, can be used to better constrain the feedback model(s). 

\section{RM estimate from the Illustris-TNG data}

\begin{figure}
\includegraphics[width=\linewidth]{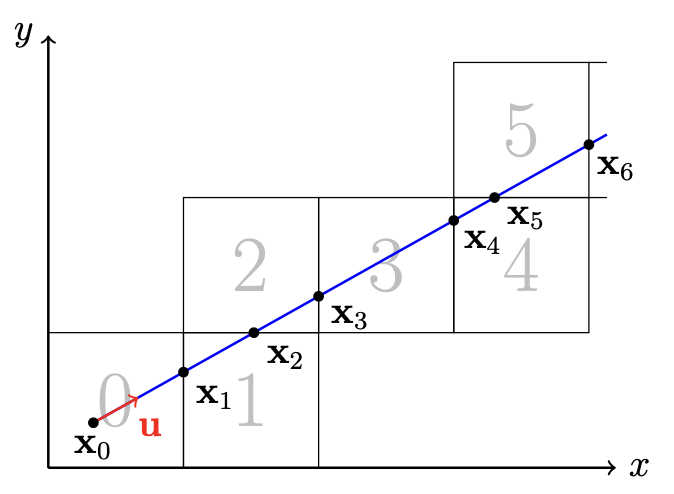}
\caption{Illustration of extension of the integration domain along the line-of-sight, using repetition of the Illustris-TNG simulation boxes.}
\label{fig:boxes}
\end{figure}

We apply Eq. (\ref{eq:RM1}) to the Illustris-TNG data to calculate the RM of extragalactic sources accumulated in the redshift interval up to $z\sim 1.5$. This redshift range spans several time snapshots of the Illustris-TNG. The box size is $L_{box}\simeq 100$~cMpc, which corresponds to $z\simeq 0.023$. To extend the integration range, we use the periodicity of the boundary conditions of the Illustris-TNG simulation and repeat the box while continuing the line-of-sight, as illustrated in Fig. \ref{fig:boxes}. We fix all crossing points $x_i$ when adding the i-th box to ensure no random shift in the position of the line while crossing multiple boxes since the generated orientation is totally random. Hence, the line-of-sight is unlikely to encounter many repeated structures. We replace the snapshot at fixed redshift with the next redshift snapshot as soon as the co-moving distance along the integration line becomes larger than the co-moving distance corresponding to the snapshot redshift.

\begin{figure}
\includegraphics[width=\linewidth]{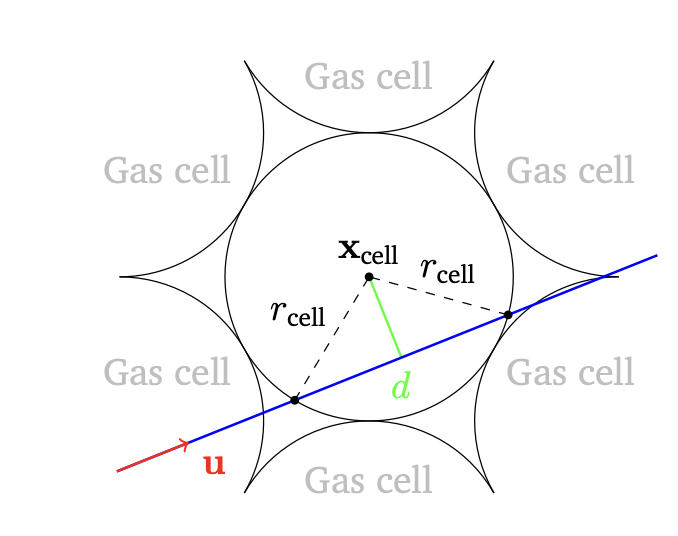}
\caption{Illustration of the passage of the line-of-sight through an Illustris-TNG gas cell.}
\label{fig:gas_cell}
\end{figure}

In each simulation box, the line-of-sight intersects gas cells of variable comoving gas density $\tilde \rho$, mass $m$, free electron abundance $\chi_e$, comoving magnetic field $\tilde B$. The comoving electron density in the cell is $\tilde n_e=\chi_e X_H \tilde\rho/m_p$, where $X_H$ is the hydrogen mass fraction and $m_p$ is the proton mass. 
%\textcolor{red}{AN: why do we ignore helium?} \textcolor{blue}{JE: They define $n_e$ as the electron number density with respect to the total hydrogen number density $n_H = X_H \tilde\rho/m_p$ so $\chi_e$ considers the contribution from both hydrogen and helium (see also eq. 31 in https://arxiv.org/pdf/astro-ph/9509107.pdf)}. 
The comoving cell size can be calculated assuming spherical symmetry $r_{cell}=\left(3m/(4\pi\tilde\rho)\right)^{1/3}$. We calculate the distance $d$ between the line-of-sight and the center of the gas cell to see if the line-of-sight  intersects the cell: $d<r_{cell}$ (see Fig. \ref{fig:gas_cell}). 

For each $i$-th cell intersecting the line-of-sight, we extract the components of the comoving magnetic field $\mathbf{\tilde B}$ and calculate  its line-of-sight component $\tilde B_{||}$. We also extract the electron abundance $\chi_e$ and the gas density $\tilde\rho$ to calculate the expression under  the RM integral (\ref{eq:RM1})
\begin{equation}
f_i(z_i)=\frac{(1+z_i)^2\tilde n_{e,i}\tilde B_{||,i}}{\sqrt{\Omega_M(1+z_i)^3+\Omega_\Lambda}}
\end{equation}
We then define $N_z=10^5$ redshift grid points between $z=0$ and $z=1.5$,  homogeneously spaced by $\Delta z=1.5\times 10^{-5}$, and interpolate the values of $f_i$ on this grid. We then calculate the RM along each line-of-sight as a sum: $RM=\sum_i f_i\Delta z$.

\begin{figure*}
\includegraphics[width=0.9\linewidth]{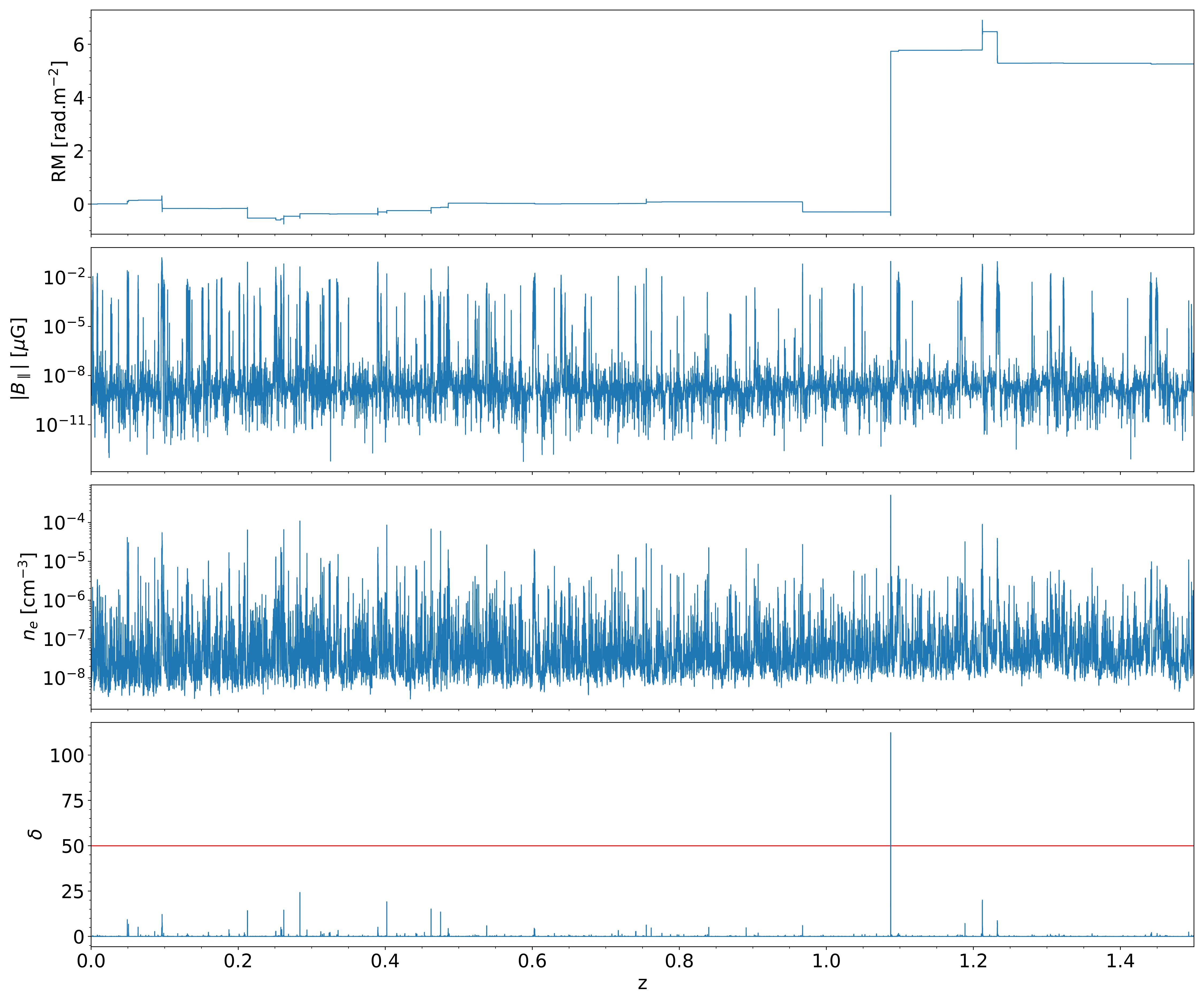}
\caption{From top to bottom: The RM, parallel component of the magnetic field, electron density and overdensity along one line-of-sight.}
\label{fig:los}
\end{figure*}

Fig. \ref{fig:los} shows the result of calculation for one example line-of sight. Large increments of the RM are typically found when the line-of-sight passes closer to a  galaxy and possibly intersects one of the magnetized bubbles produced by the baryonic outflows. The probability of such crossing increases with the redshift and, sooner or later, all lines-of-sight would cross at least one magnetized bubble. 

%%%%%%%%%%%%%%%%%%%%%%%%%%
%%%%%%%%%%%%%%%%%%%%%%%%%%

\begin{figure}
\includegraphics[width=\linewidth]{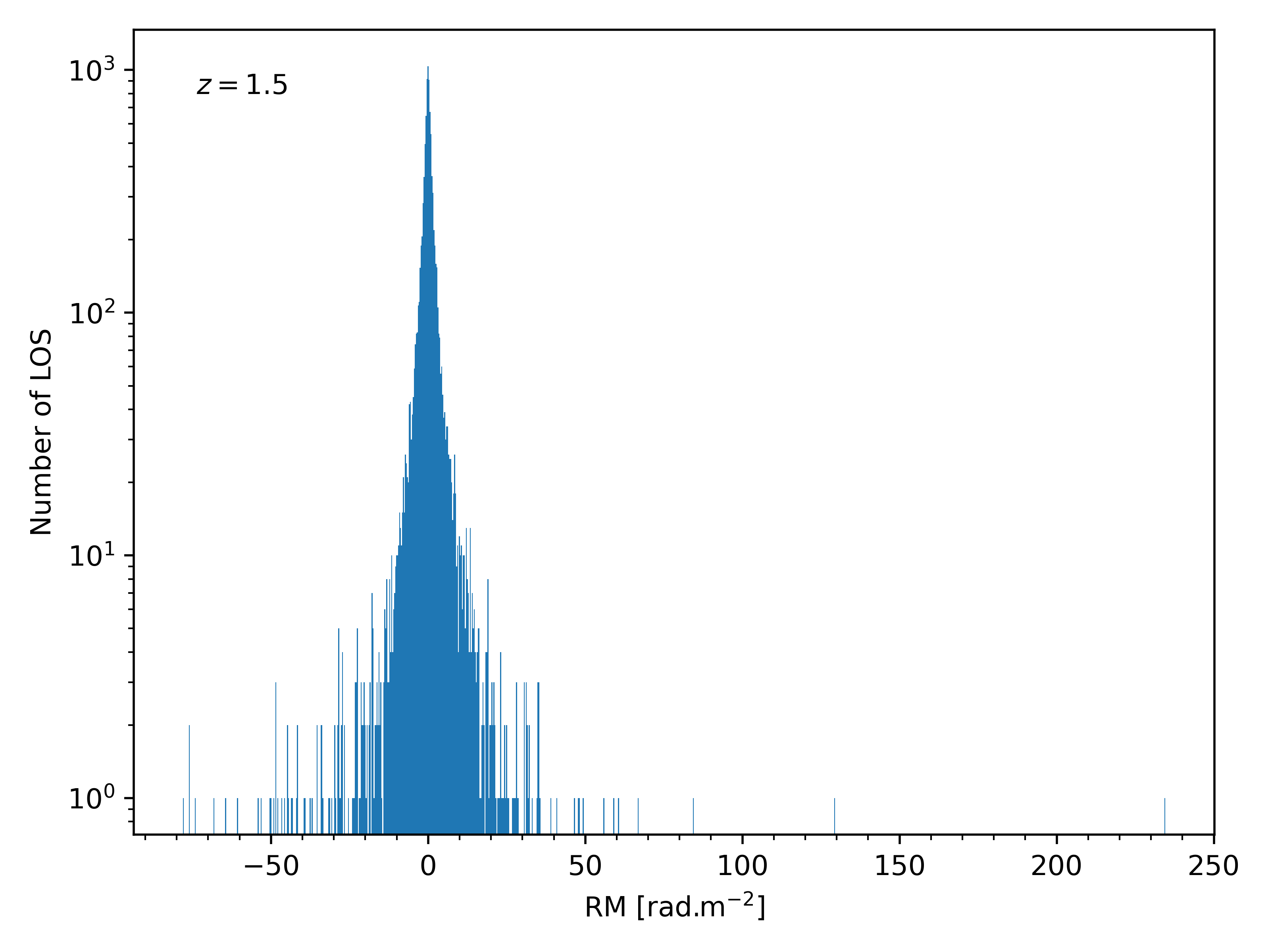}
\caption{Histogram of the RM values at $z=1.5$ for a sample of $10^4$ lines-of-sight with $\delta < 50$.}
\label{fig:RM_histogram}
\end{figure}

Fig. \ref{fig:RM_histogram} shows a histogram of the RM values for a sample of $10^4$ randomly generated lines-of-sight. 
%The mean of the distribution is $\left< RM\right>=0.05$ rad/m$^2$, the standard deviation is $\sigma_{RM}=4.7$~ rad/m$^2$. 
One can see a number of outliers characterised by large RM that correspond to the cases when the line-of-sight  occasionally crosses a galaxy or galaxy cluster. The analysis of \cite{2023MNRAS.518.2273C} has shown that the RM sample of LOTSS mostly does not contain lines-of-sight passing through large overdensities of the LSS. To account for this fact in generation of predictions for the RM from numerical simulations, \cite{2023MNRAS.518.2273C} have found that the closest approach of the lines-of-sight to the sources used in the LOTSS RM dataset corresponds to the gas cells with overdensities  $\delta=\tilde \rho/\tilde\rho_{crit}\sim 50$ ($\tilde\rho_{crit}$ is the critical density of the Universe). To account for the absence of the lines-of-sight passing through larger overdensities, \cite{2023MNRAS.518.2273C} have removed cells with densities larger than this limit from the integration along the lines-of-sight through their simulation volumes. We adopt the same approach and remove gas cells with overdensities larger than the threshold $\delta_{thr}=50$ from the integral of Eq. (\ref{eq:RM1}).  

%%%%%%%%%%%%%%%%%%%%%%%%%%%%%%%%%%%%
%%%%%%%%%%%%%%%%%%%%%%%%%%%%%%%%%%%%

%%%%%%%%%%%%%%%%%%%%%%%%%%%%%%%%%%%%
\section{Results}
%%%%%%%%%%%%%%%%%%%%%%%%%%%%%%%%%%%%

%%%%%%%%%%%%%%%%%%%%%%%%%%%%%%%%%%%%
\begin{figure}
\includegraphics[width=\linewidth]{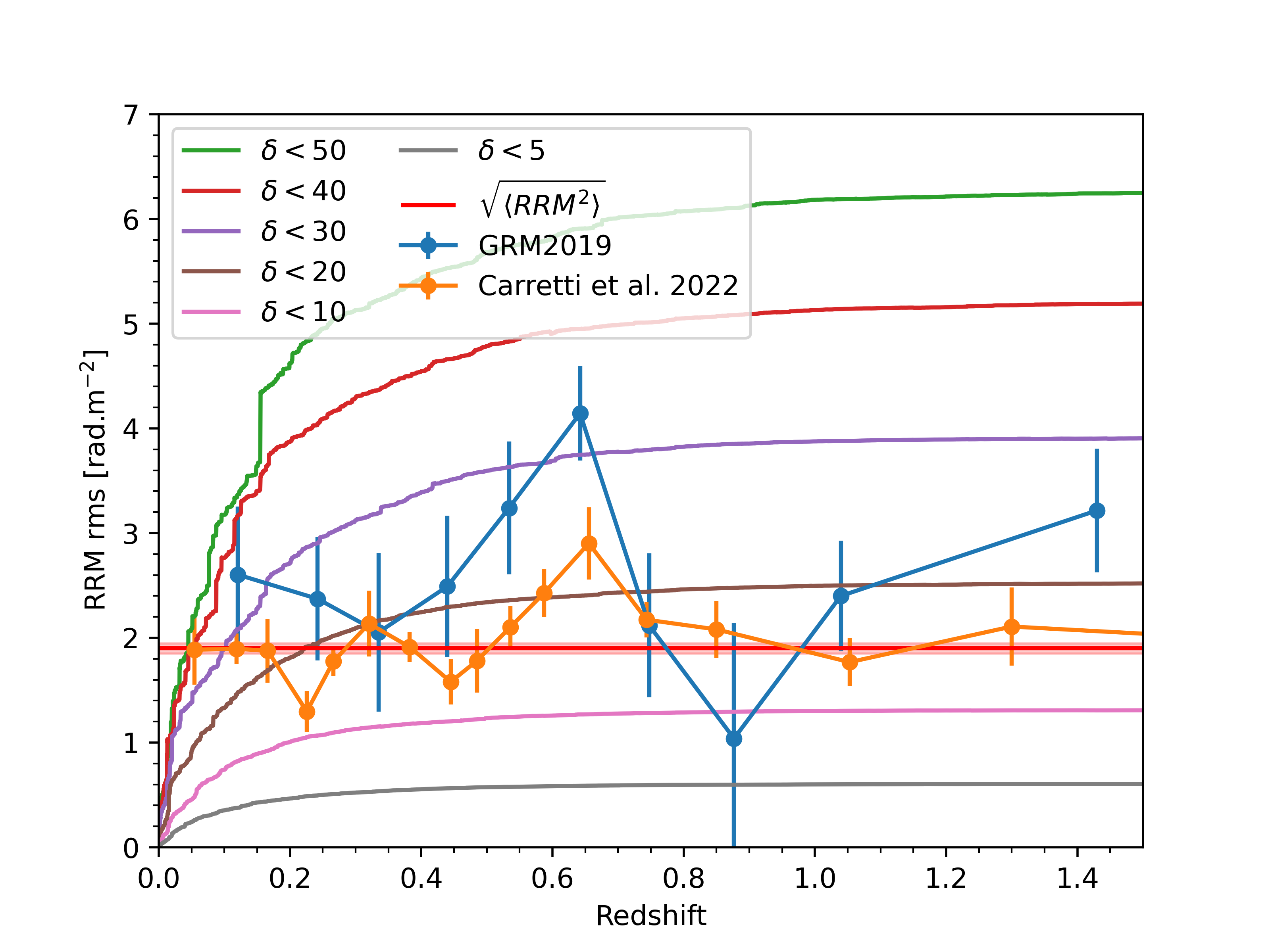}
\caption{Evolution of the RMS of the RM with the redshift. Measurement of the RMS of the RRM derived from the LOFAR data \citep{2023MNRAS.518.2273C} based on the GRM model of \citet{2022A&A...657A..43H}, is shown in individual redshift bins (orange data points) and a constant $\sqrt{\langle RRM \rangle}$ fit (red horizontal band). Blue data points show the RMS of the RRM calculated using the GRM model of \citet{2020A&A...633A.150H}.}
\label{fig:RM_redshift}
\end{figure}
%%%%%%%%%%%%%%%%%%%%%%%%%%%%%%%%%%%%

Fig. \ref{fig:RM_redshift} shows the redshift evolution of the root-mean-square (RMS) of the RM calculated for a sample of $10^4$ lines-of-sight through the Illustris-TNG simulation. The RM quickly accumulates in the redshift range up to $z\sim 1$ and stabilises at larger redshifts. The main reason for the fast growth of the RM at small redshifts is the presence of the magnetized bubbles driven by the baryonic feedback \citep{2023MNRAS.519.4030A}. These bubbles are most actively driven by the star formation and AGN activity that peaks at somewhat higher redshifts. Stabilisation of the RM at redshifts $z>1$ indicates that the baryonic feedback driven bubbles reach their maximal volume filling factor by the redshift $z\sim 1$. The seed cosmological magnetic field assumed in Illustris-TNG is at the level $10^{-14}$~G and its contribution to the RM is negligible compared to that of the outflow-driven magnetic field. 

Fig. \ref{fig:RM_redshift} also shows a comparison of the estimate of the RM from Illustris-TNG with the Residual Rotation Measure (RRM) extracted from the LOFAR data by \citet{2023MNRAS.518.2273C}. The overall RM measured by LOFAR has contributions from the Galactic RM (GRM) accumulated during the signal passage through the Milky Way galaxy, the RM accumulated during the signal passage through the intergalactic space, $RM_{IGM}$ (of interest in this work) as well as the RM accumulated at the extragalactic radio source, RM$_{local}$:
\begin{equation}
RM=GRM+RM_{IGM}+RM_{local}+RM_{noise}
\end{equation}
There is also an additional "noise" component present because of the limited precision of the measurements.
The "residual" RM is obtained after subtraction of the GRM model from the total RM:
\begin{equation}
RRM=RM-GRM
\end{equation}
By construction, the RRM includes the RM local to the radio source and the measurement noise, in addition to the RM accumulated in the IGM. Due to the presence of the noise and  source components, the RMS of the RRM is expected to be larger that the RMS of the intergalactic RM component only. Thus, the measurement of the RMS of the RRM derived by \citet{2023MNRAS.518.2273C} 
\begin{equation}
\sqrt{\left<RRM^2\right>}=1.90\pm 0.05\mbox{ rad/m}^2
\end{equation}
should be considered as an upper bound on the RMS of the intergalactic RM:
\begin{equation}
RM_{IGM}<\sqrt{\left<RRM^2\right>}.
\end{equation}
This upper bound is shown in Fig. \ref{fig:RM_redshift} by a horizontal line. 
%\textcolor{red}{AN: can you add the horizontal line (with an error band) to the figure?} \textcolor{blue}{JE: Done. Should we also remove the data points from Caretti to simplify the plot ?}.

One can see that the LOFAR upper bound on the intergalactic RM  is in contradiction with the estimate of the RM extracted from the Illustris-TNG model, already starting from the redshift $z\simeq 0.01$. This conclusion is valid if the cut $\delta<50$ is imposed on the gas cells included in the RM integral, as explained in the previous section. At larger redshifts, $z\sim 1$, the discrepancy between the LOTSS upper limit and Illustris-TNG estimate reaches a factor of $3$. Fig. \ref{fig:RM_redshift} shows that the model predictions can be made consistent with the data only if all the gas cells with overdensity in excess of $\delta_{thr} \simeq 15$ are removed from the RM integral of Eq. (\ref{eq:RM1}). This would efficiently remove parts of the lines-of-sight passing through filaments of the LSS, which would not correspond to the realistic lines-of-sight from redshifts $z\gtrsim 1$ in the LOTSS sample that are unlikely to completely avoid filaments.

%%%%%%%%%%%%%%%%%%%%%%%%%%%%%%%
\section{Discussion}
%%%%%%%%%%%%%%%%%%%%%%%%%%%%%%%

Contradiction between the estimate of the intergalactic RM from Illustris-TNG and LOFAR data shows that the model of magnetisation of the intergalactic medium implemented in Illustris-TNG needs to be revised. The modeling shows that  the RM along the lines-of-sight mostly accumulates in  the magnetised bubbles around galaxies. This contrasts with the models considered by \citet{2023MNRAS.518.2273C} in which the intergalactic RM was accumulated in filaments occupied by the compressed primordial magnetic field. This is explained by much stronger assumed initial field strength ($\sim 10^{-10}$~G in \citet{2023MNRAS.518.2273C} vs. $10^{-14}$~G in Illustris-TNG). Moreover, the Illustris-TNG baryonic feedback model prediction for the RM is dramatically different from the estimate of \citet{2023MNRAS.518.2273C}, where the "astrophysical" model of the magnetic field spread by galactic outflows failed to even produce the RM reaching the level observed by LOFAR. This perhaps demonstrates the wideness of the uncertainty range of the feedback parameters. Another possibility is that phenomenological model of the outflows is right, but it is the numerical scheme used for solution of the MHD equations that over-estimates the strength of the fields in the outflows \citep{2016MNRAS.455...51H,2016MNRAS.463..477M}.

\citet{2023MNRAS.519.5723O,2023MNRAS.518.2273C} have noticed that the GRM estimates of \citet{2022A&A...657A..43H} are biased toward the RM measurements by LOFAR at the locations of sources from LOTSS, because the RM data sample used by \citet{2022A&A...657A..43H} includes the LOFAR data of superior quality compared to the bulk of the RM data used for the GRM modelling. This led to an under-estimate of the RRM. To overcome this difficulty, \citet{2023MNRAS.519.5723O,2023MNRAS.518.2273C} used the GRM model smoothed over one degree on the sky. To verify the estimate of the RRM of \citet{2023MNRAS.519.5723O,2023MNRAS.518.2273C}, we compare the RRM estimate of \citet{2023MNRAS.519.5723O,2023MNRAS.518.2273C} 
 with that derived from the same LOFAR data, but using the GRM model of \citet{2020A&A...633A.150H} that did not include LOFAR data. The result, shown by the blue data points in Fig. \ref{fig:RM_redshift}, shows that the two estimates are consistent with each other. Tension between Illustris-TNG prediction and the RRM data is thus unlikely to be due to the under-estimate of the RRM in LOFAR data.

Even though the comparison of the Illustris-TNG model points to the need of revision of the model, it illustrates that the baryonic feedback field may readily saturate the bound on the intergalactic RM from LOTSS.  This suggests that, contrary to the claim of \citet{2023MNRAS.518.2273C}, presence of the $40-80$~nG field found in the LSS filaments does not necessarily favour a model of the primordial stochastic magnetic field amplification in the filaments. Fields spread by galactic outflows with somewhat reduced volume filling factor, compared to Illustris-TNG model, may well be responsible for the fields found in the filaments.  

One more important implication of our result concerns the possibility of measurement of cosmological relic magnetic fields in the present-day Universe using the methods of \gr\ astronomy \citep{Bondarenko:2021fnn}. In the Illustris-TNG model, the magnetised outflows from the galaxies are not strong enough to spread into the voids of the LSS with large volume filling factor \citep{2023MNRAS.519.4030A}. Our result shows that the Illustris-TNG model over-estimates the strength of the outflows and suggests that the volume filling factor of the outflows is still smaller. This supports the idea  that the large-scale magnetic fields found in the voids \citep{2010Sci...328...73N,2023A&A...670A.145A} are of cosmological origin. 

\section*{Acknowledgements} We would like to thank F.Vazza, A.Boyarsky, K.Bondarenko and A.Sokolenko for useful comments on the text.

\bibliographystyle{aa}
\bibliography{refs.bib}

\end{document}